\documentclass{aa}  

\usepackage{natbib}
\usepackage{graphicx}
\usepackage{txfonts}
\usepackage{hyperref}
\usepackage{mathtools}
\usepackage{subcaption}
\usepackage{balance}

\begin{document} 

  \title{Detecting transit signatures of exoplanetary rings using SOAP3.0}
  \author{B. Akinsanmi 
          \inst{1,2,4}, M. Oshagh\inst{1,3}, N.C. Santos\inst{1,2}, and S.C.C. Barros\inst{1}       
          }

  \institute{  Instituto de Astrofísica e Ciências do Espaço, Universidade do Porto, CAUP, Rua das Estrelas, 4150-762 Porto, Portugal.\\
  \email{tunde.akinsanmi@astro.up.pt}
 \and
 Departamento de Física e Astronomia, Faculdade de Ciências, Universidade do Porto, Rua do Campo Alegre, 4169-007 Porto, Portugal.
 \and  Institut für Astrophysik, Georg-August-Universität Göttingen, Friedrich-Hund-Platz 1, 37077 Göttingen, Germany.
 \and National Space Research and Development Agency. Airport Road, Abuja, Nigeria.
 }


  \abstract
   {It is theoretically possible for rings to have formed around extrasolar planets in a similar way to that in which they formed around the giant planets in our solar system. However, no such rings have been detected to date.}
   {We aim to test the possibility of detecting rings around exoplanets by investigating the photometric and spectroscopic ring signatures in high-precision transit signals.}
   {The photometric and spectroscopic transit signals of a ringed planet is expected to show deviations from that of a spherical planet. We used these deviations to quantify the detectability of rings. We present SOAP3.0 which is a numerical tool to simulate ringed planet transits and measure ring detectability based on amplitudes of the residuals between the ringed planet signal and best fit ringless model.}
   {We find that it is possible to detect the photometric and spectroscopic signature of near edge-on rings especially around planets with high impact parameter. Time resolution $\leq$7\,mins is required for the photometric detection, while 15 mins is sufficient for the spectroscopic detection. We also show that future instruments like \textit{CHEOPS} and \textit{ESPRESSO}, with precisions that allow ring signatures to be well above their noise-level, present good prospects for detecting rings.}
   {}

\titlerunning{SOAP3.0}
\authorrunning{B. Akinsanmi et al.}
\maketitle


\section{Introduction}
Planetary transits offer very valuable information about planets which are not accessible through other planet detection techniques. When planets transit their host star, they obscure part of the stellar light. Photometrically, a dimming of the stellar light is observed thereby producing a light-curve. Spectroscopically, some of the radial velocity (RV) components of the rotating star is blocked causing an anomaly (line profile asymmetry) referred to as the Rossiter-McLauglin (RM) effect \citep{rossiter,mclaughlin}. 

Photometric transits have been used to detect a large number of the exoplanets known today. Furthermore, the photometric and spectroscopic transit techniques allow for the characterisation of properties such as planetary radius, orbital inclination and velocity of stellar rotation amongst others. These transit techniques have been applied in the detection of multi-planetary systems \citep{gillon17}, study of planet oblateness \citep{caterwinnb,carterwinna,zhu}, investigation of exoplanetary atmospheres using transmission and occultation spectroscopy \citep{chab,deming,madhu} and also in the measurement of spin-orbit misalignment \citep{sanchis,addison}.

Planetary rings are unique features in our solar system yet to be detected around extrasolar planets. The giants planets in our solar system all have rings with different radial extents and particle size. Just as the planets in our solar system motivated the search for exoplanets, the rings of the giant planets have raised questions on the existence of rings around exoplanets \citep{brown,sch11,ken}. The detection of exoplanetary rings would have tremendous astronomical consequences and could usher a paradigm shift in our understanding of planetary formation and evolution. For instance, detecting rings around short-period giant planets, similar or dissimilar to those in our solar system, would require explanations as to how they might have formed.

\citet{sch11} investigated the nature of the rings around close-in planets proposing the possibility of rock-like rings instead of the icy rings around our solar system planets. They also emphasised the possibility of detecting rings at semi-major axes larger than 0.1 AU. \citet{barnes} showed that photometric precision of $100-300\,$ppm with 15 minute time resolution would suffice for the detection of Saturn-like rings around an exoplanet. \citet{zuluaga} presented a large-scale photometric survey method to identify ringed planet candidates. The method uses the anomalously large transit depth and anomalous estimation of transit derived stellar density to probe the presence of a ring. The search for rings in photometric data has also been performed. \citet{heising} searched for ringed planets around 21 short-period planets in \textit{Kepler} photometry and found no evidence for rings. \citet{aizawa} also searched for rings in \textit{Kepler} photometry around 89 long-period planet candidates that exhibit transit-like signals. They found a planet candidate whose transit signal could be explained by one of three scenarios:  the presence of a planetary ring, a circumstellar disc or a hierarchical triple.  \citet{lecav} searched for transit signature of satellites and rings around the long-period planet CoRoT-9b and excluded the presence of both bodies around the planet.
In addition to photometric detection of rings, \citet{ohta} studied complementary spectroscopic detection. They concluded that rings could be detected with radial velocity precision of 1\,m/s if not viewed close to edge-on. 

In this paper, we present a new tool, "SOAP3.0", modified and developed to simulate the photometric and spectroscopic transit of a ringed planet. It delivers the expected light-curve and RM signal for different ring orientations. 

In Sect. 2, we describe the developments that lead to SOAP3.0, explain its input and output parameters and make comparisons of the results with those in literature. In Sect. 3 we apply the tool to investigate the detectability of different ring orientations using high-precision transit photometry and radial velocity. In Sect. 4, we discuss factors that can influence the ringed planet signal and ring signature and also the prospects of upcoming instruments for detecting rings. We draw conclusions in Sect. 5.

\begin{figure}
\centering
\includegraphics[width=1\linewidth]{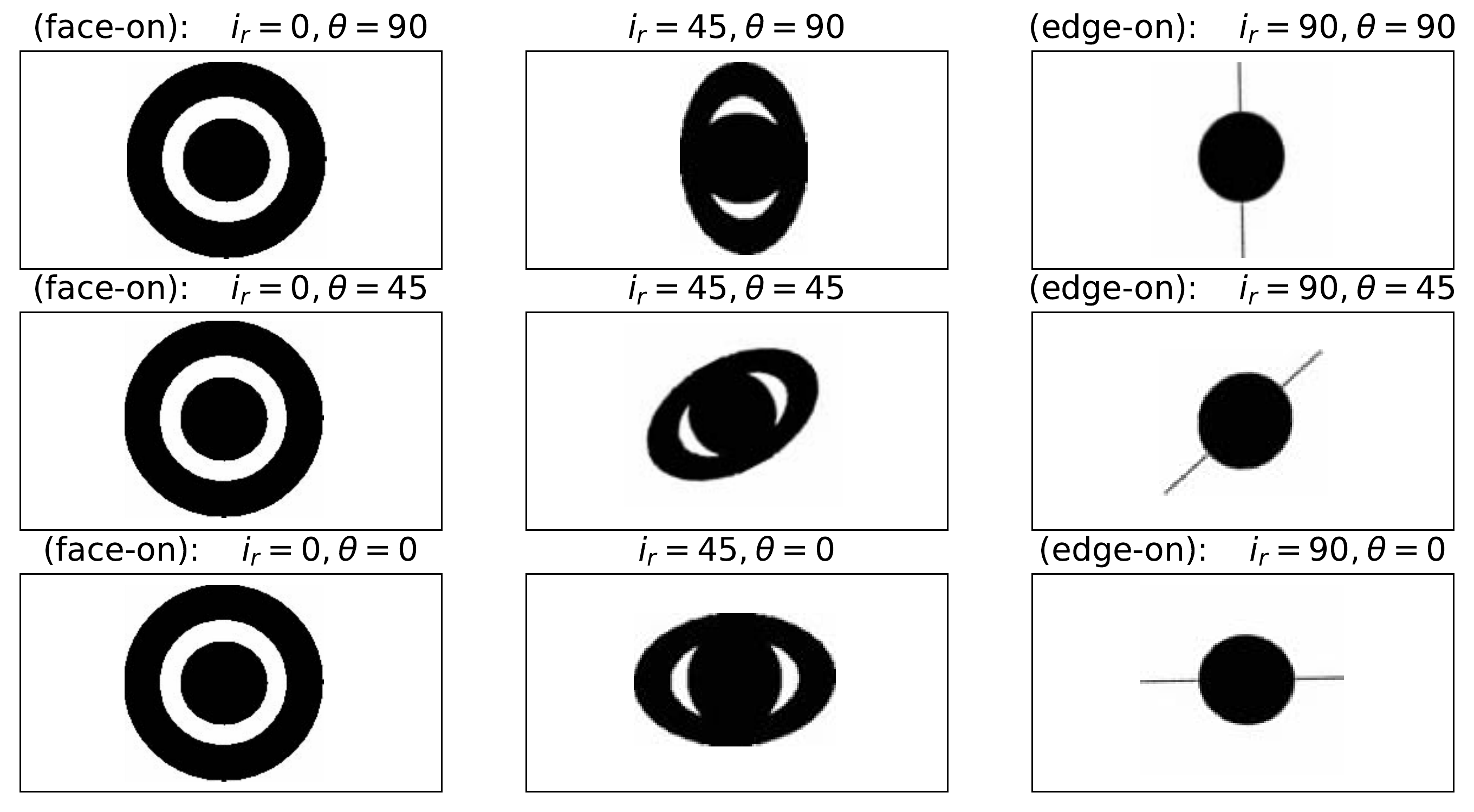}
\caption{Illustration of a planet with 9 different ring orientations defined by $i_{r}$ and $\theta$.}
\label{illus}
\end{figure}

\section{ SOAP3.0 - Ringed planet transit}
SOAP3.0 is a numerical tool developed as a modification to the planet transit tool SOAP2.0-T by \citet{oshagh-2016}. SOAP2.0-T is an adaptation of SOAP \citep{boisse} and SOAP2.0 \citep{dumus} in order to simulate the photometric and radial velocity variations of a planet transiting a rotating spotted star. It generates the transit light-curve and Rossiter McLauglin (RM) signal due to the transiting planet. Detailed descriptions of these tools can be found in \citet{boisse,oshagh-soapt} and \citet{dumus}.
    
We developed SOAP3.0 to simulate the transit light-curve, RM signal and the induced anomalies in these signals due to the transit of the ringed planet. This tool is also capable of generating the signal variations due to occultation of stellar active regions by the ringed planet but this is not thoroughly investigated here as it is not the the focus of this paper.
    
To add the effect of rings to the computation, we assumed that rings are circular, geometrically thin and completely opaque. The rings lie beyond the planet's radius and blocks stellar light in the same way as the planet but only between inner and outer radii defined by $R_{in}$ and $R_{out}$. The ring orientation is defined by two angles: $i_{r}$, the inclination of the ring plane with respect to the sky plane ($0^\mathrm{o}$ and $90^\mathrm{o}$ for face-on and edge-on projections respectively)  and $\theta$, the tilt of the ring plane to the planet's orbital plane ($0^\mathrm{o}$ and $90^\mathrm{o}$ for ring projection parallel and perpendicular to orbital plane respectively). Figure \ref{illus} illustrates different ring orientations for a planet. It is assumed that the ring maintains the same orientation throughout the transit phase\footnote{Transit duration is much shorter than orbital period so change in ring orientation during transit is negligible.}. 


\subsection{Input parameters}
SOAP3.0 is supplied with input parameters through a configuration file "config.cfg". Inputs are as defined in \citet{oshagh-soapt}. Here we describe the relevant input parameters for ringed planet transit.

For a spherical ringless planet transit, the input parameters are: the radius of the planet $R_{p}$ (in units of stellar radii $R_{\ast}$), the semi-major axis $a$ (in units of stellar radii), planet orbital inclination $i_{p}$, and the planet's orbital period $P_{p}$ (in days) calculated from $a$ and stellar mass using Kepler's third law. The impact parameter $b$ can be calculated using $b=a\cos i_{p}$. Additional inputs include: periastron passage time $t_{0}$, eccentricity $e$, argument of periastron $\omega$ and the planet's initial phase $\psi_{0}$.

For the ring, the input parameters to the code are: inner ring radius $R_{in}$ and outer ring radius $R_{out}$ both in units of planet radii ($R_{p}$). Also the two orientation angles: inclination $i_{r}$ and tilt $\theta$ both in degrees.

\subsection{Outputs}
The output of the code gives the Flux, RV, BIS (bisector span), and FWHM\footnote {FWHM of cross correlation function} variations and can be plotted as a function of stellar rotation phase, orbital phase or time.  Details of how these variations are computed can be seen in \citet{dumus,oshagh-soapt,boisse}.

\begin{figure}
\centering
\includegraphics[width=1\linewidth]{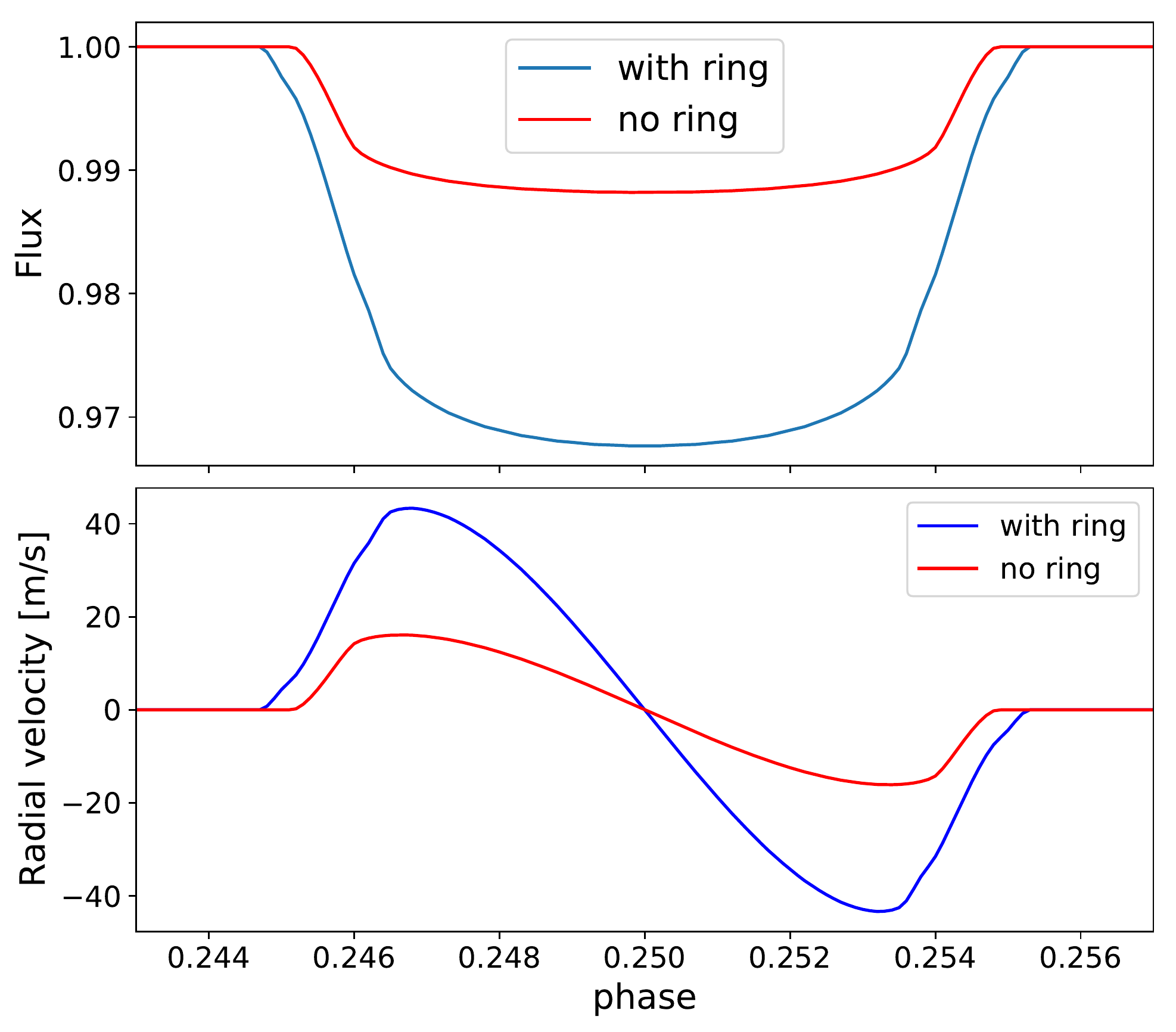}
\caption{SOAP3.0 Light-curve (top) and RM signal (bottom) for a transiting planet without a ring (red) and with a ring (blue) at face-on orientation ($i_{r},\theta=0,0$). The ringed planet produces a longer transit of 6.38\,hrs compared to 5.85\,hrs for the ringless planet.}
\label{ringnoring}
\end{figure}

\begin{figure*}[ht!]
        \centering
        \includegraphics[width=1\linewidth]{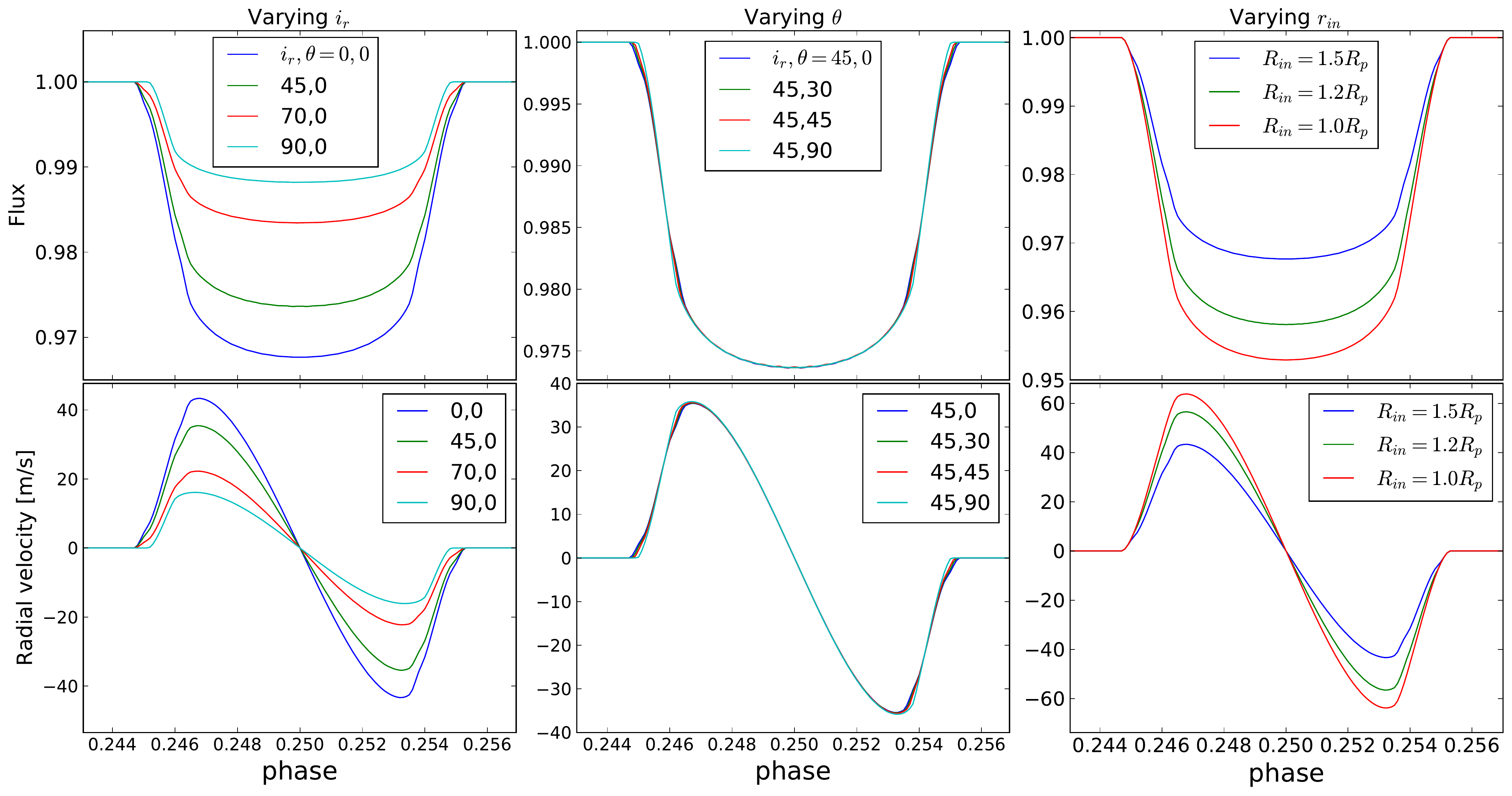}
        \caption{Effect of varying ring parameters on the light-curve and RM signals. First column shows how the light curve and RM signal varies for different values of $i_{r}$. Second column shows signal variation for different values of $\theta$ when $i_{r}=45$. Third column shows the effect of varying the gap between planet surface and ring area by varying $R_{in}$.}
        \label{varying}
\end{figure*}

\subsection{Physical limits on ring input parameters}
The existence of rings around close-in planets are not known but dynamical arguments suggest that detectable rings are possible. \citet{sch11} showed that exoplanets with $a$\,$>$\,0.1\,AU could host silicate rings and if optically thick, the rings would have long life time of up to $10^9$ years. Rings around these planets would have to be in the stable regions of the planet's Hill radius and also within the Roche radius \citep{santos}.

\subsection{SOAP3.0 transit signals: Light-curve and RM signal}
To illustrate the output of SOAP3.0, we simulated a planetary transit with $\sim$200\,s (3.5\,mins) time-sampling using fiducial values in Table \ref{simul}. The planet is a Jovian-like planet with a semi-major axis of 0.16\,AU ($36.08\,R_{\ast}$) assumed to follow a circular orbit around a quiet solar-like star with $T_{\mathrm{eff}}$=5778\,K. The velocity of stellar rotation $\nu\sin i_{\ast}$ is assumed to be edge-on (rotation axis parallel to sky plane). Stellar quadratic limb darkening coefficients given by \citet{claret} are used for the star described.

Figure \ref{ringnoring} shows the transit light-curve and RM signal for a spherical ringless planet and for the same planet having the ring parameters in Table \ref{simul}. The ring orientation is face-on (i.e. $i_{r},\theta=0,0$). It is seen that the ringed planet produces a deeper photometric transit and greater RM amplitude than the ringless planet. This is due to the additional stellar disc area covered by the ring. Also since the ring increases the projected radial extent of the planet, it causes a longer transit duration (6.38\,hrs). This light-curve and RM signal can however be easily produced by a planet with a larger radius.

   \begin{table}[hb]
        \centering
        \caption{Simulation parameters selected to satisfy the physical ring limits}
        \label{simul}

        \begin{tabular}{l l l}
                \hline\hline
                Parameter & Value  &Description\\ 
                \hline
                $R_{\ast}$ [$R_{\odot}$]   & 1.0           &  Stellar radius  \\
                $u_{1}$, $u_{2}$        & 0.29, 0.34        &  Limb darkening coefficients\\
                $\nu\sin i_{\ast}$ [km/s]       & 2    &Stellar rotation velocity         \\
                \hline
                $a$ [$R_{\ast}$]                & 36.08       &Semi-major axis   \\
                $P$ [days]                      & 25           & Orbital period  \\
                $i_{p}$ [$^\mathrm{o}$] & $90$ $(b=0)$  &Orbital inclination\\
                $R_{p}$ $[R_{\ast}]$            & $0.1$   &Planetary radius        \\
                $\lambda$ [$^\mathrm{o}$]       & $0$      &Spin-orbit misalignment angle \\
                \hline
                $R_{in}$ [$R_{p}$]              & $1.5$      &Ring inner radius    \\
                $R_{out}$ [$R_{p}$]             & $2.0$       &Ring outer radius        \\
                $i_{r}$ [$^\mathrm{o}$]          & [0,90]      & Ring inclination    \\
                $\theta$ [$^\mathrm{o}$]         & [0,90]  &Ring tilt    \\
        \hline                          
        \end{tabular}
   \end{table}

A feature more indicative of the presence of ring is the anomaly seen when the ring's outer edge, inner edge and planet edge contact the stellar disc at ingress and egress phases. These anomalies manifest as wiggles in the transit light-curve and RM signal during ingress and egress as seen in Fig. \ref{ringnoring}. Therefore, the detection and identification of exoplanetary rings depend on the ability to detect and measure these wiggles as will be seen in Sect. 3.

The signals from the occultation of a stellar spot by the ringed planet is shown in Appendix A. A detailed impact of stellar spots will be duly explored in a future work.

   \begin{figure*}[!th]
\centering
\includegraphics[width=1\linewidth]{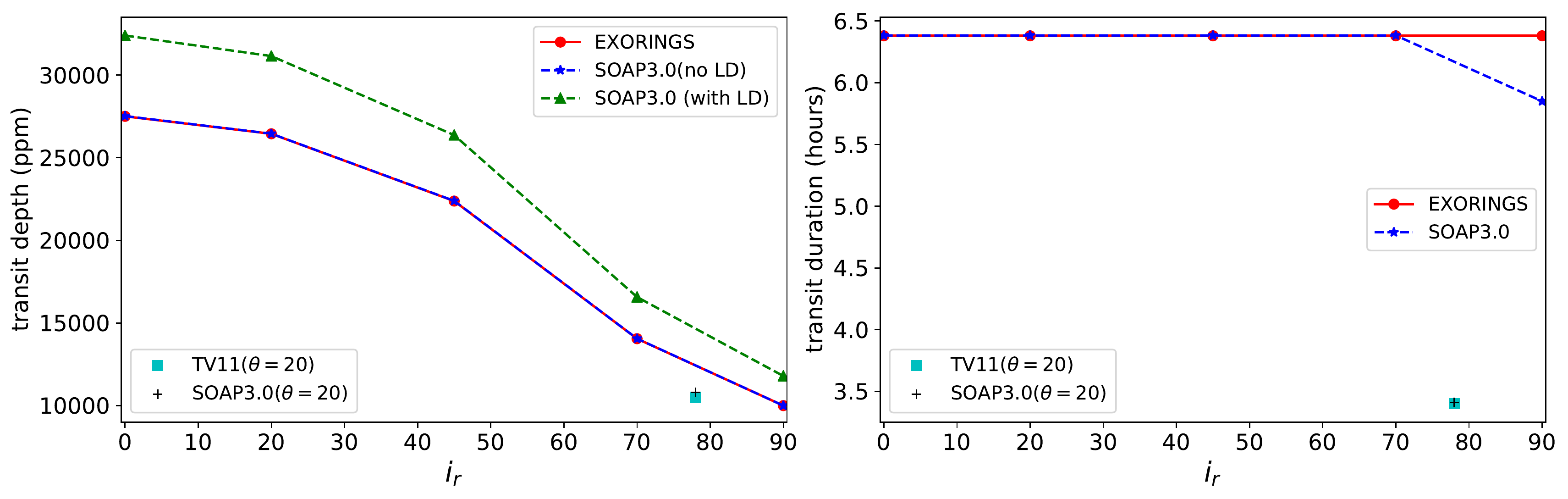}
\caption{Comparisons of SOAP3.0 results with those from EXORINGS and \citet{tsunski} [TV11]. Transit depth (left pane) and transit duration (right pane) as a function of $i_{r}$ for $\theta=0$ computed using \textit{SOAP3.0} and \textit{EXORINGS}. Also comparison of SOAP3.0 with TV11 for $i_{r},\theta=78,20$. Green triangles are points from SOAP3.0 using solar limb darkening, blue asterisks are points from SOAP3.0 without limb darkening and the red circles are points from \textit{EXORINGS}. Cyan squares and black crosses are the points from TV11 and SOAP3.0 respectively for $i_{r},\theta=78,20$.}
\label{compare}
\end{figure*}

For the same planet, the transit signal varies as the parameters of the ring changes. We illustrate the effect of ring parameter changes on transit light-curve and RM signal in Fig. \ref{varying}. As seen in its first column, varying $i_{r}$ from face-on ($i_{r}=0^\mathrm{o}$) up to edge-on ($i_{r}=90^\mathrm{o}$) causes the transit signals to decrease in amplitude due to the reduction in ring projected area with $i_{r}$. At edge on, the light-curve and RM signal of the ringed planet appears indistinguishable from that of the ringless planet since the thickness of ring is negligible and does not block any stellar flux. The second column shows the signals when $i_{r}$ is kept at $45^\mathrm{o}$ and $\theta$ is varied from $0^\mathrm{o}$ to $90^\mathrm{o}$. It is seen that the signals do not vary very much with $\theta$, its most visible effect is to slightly reduce the transit duration as it approaches $90^\mathrm{o}$. The third column of Fig. \ref{varying} shows the effect of changing the inner radius of the ring while the outer radius remains at the constant value in Table \ref{simul}. The value of $R_{in}$ determines the size of the gap between planet surface and ring and it is seen that as the gap size reduces the signal amplitude increases due to increasing ring area.

The plots in Fig. \ref{varying} has shown that varying ring parameters mostly affects the amplitude and to a much lesser extent the duration of the transit signals. More subtle variations can be noticed in the prominence of the wiggles at ingress and egress as the ring parameters change. The variation in the prominence of the wiggles with ring parameters can be used to detect and characterise the signature of the ring.

\subsection{Performance test of SOAP3.0}
SOAP3.0 is capable of producing precise transit light-curves and RM signals for spherical planet transits as shown by \citet{oshagh-soapt}. It is, however, important to validate if the inclusion of rings provide the expected output.\\

We compared the ringed planet photometric results of SOAP3.0 with those from \textit{EXORINGS} \citep{zuluaga} and \citet{tsunski}. We used Table \ref{simul} to generate mock transits of a ringed planet with both SOAP3.0 and \textit{EXORINGS} and compared their results. Then comparison with transit result shown in \citet{tsunski} was done using the same input parameters as the paper. The results of these comparisons are shown in table \ref{comparetable} of Appendix B and summarised in Fig. \ref{compare}. Limb darkening is not considered in \textit{EXORINGS} whereas SOAP3.0 uses a quadratic limb darkening law causing their results to be different. However, when limb darkening is also ignored in SOAP3.0, both tools show excellent agreement. For edge-on ($i_{r}$=$90^\mathrm{o}$) orientations where the ring should have no contribution to the transit duration and depth (since they are assumed to be infinitely thin), \textit{EXORINGS}, being analytic, calculates the transit duration inaccurately.

Comparison with \citet{tsunski} shows similar results for transit duration but not with transit depth. This is because SOAP3.0 assumes completely opaque rings while \citet{tsunski} model uses opacity $\tau$=0.5 ($\tau=[0,1]$). However, we note that since the opacity and ring area are degenerate parameters, one can compensate for the opacity by reducing the area of the ring. For instance, a ring with $\tau=0.5$ can be mimicked by an opaque ring with half the area. Indeed when we did this, the transit depth obtained is similar to \citet{tsunski}. We note here that this comparison was done based on visual inspection of the light-curve in the paper since we had no access to the code used.

We have not compared our ringed planet RM signal with that of \citet{ohta} due to difference in RV measurement technique employed in the models \citep{boue}. \citet{ohta} computes weighted mean velocity along stellar line of sight whereas we performed a Gaussian fit to a cross-correlation function \citep{boisse} as is done on stabilised spectrographs. In spite of this, visual comparison of the paper's result with that of SOAP3.0 shows that the shape of the RM signal and wiggles are very identical.

\section{Detecting ring signatures}
The ring signature in a transit signal would be the residual between the ringed planet signal and the best-fit ringless planet model \citep{barnes}. The maximum residuals indicating the ring signature should therefore be positioned around ingress and egress.
\begin{figure*}
        \centering
        \includegraphics[width=1\linewidth]{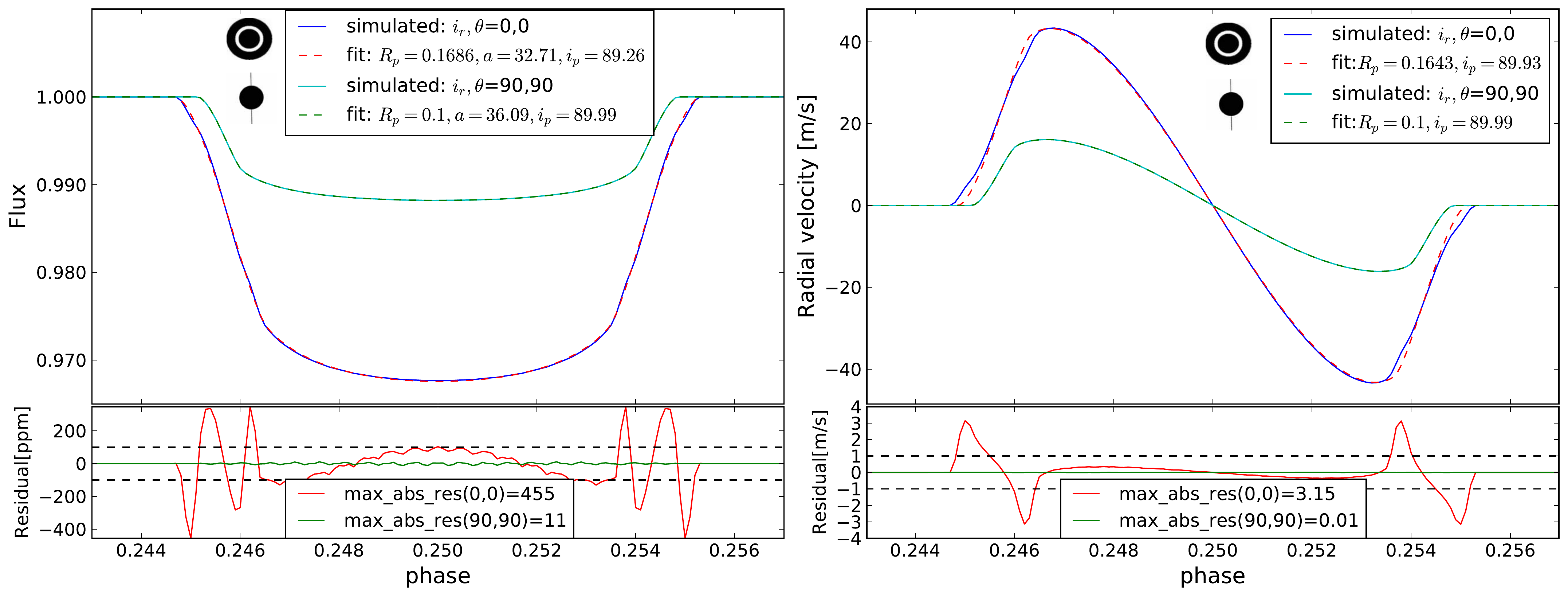}
        \caption{Ringless model fit to two ring orientations (face-on and edge-on) of the ringed planet. Left pane shows the analytical ringless fit to the two ringed planets light-curves and the respective residuals generated. Right pane shows the numerical ringless fit to the two ringed planet RM signals and the residuals. The black dashed line in residual plots show the detection limit as mentioned in text.}
        \label{fits}
\end{figure*}

We simulated a combination of nine different ring inclinations ($i_{r}$) and seven different ring tilt angles ($\theta$) to cover the range of possible ring orientations. Therefore a total of 63 ringed planet transits having different $i_{r},\theta$ combinations were simulated with SOAP3.0 using fiducial values in Table \ref{simul}. Using theoretical model of \citet{mandel02}, we then fitted the simulated ringed transit light-curves with a ringless planet model. For this fitting, we allow $R_{p}$, $a$ and $i_{p}$ to vary as free parameters while the limb darkening coefficients (LDC) $u_{1}$ and $u_{2}$ are fixed and assumed to be known a priori (e.g. from \citealt{claret} or \citealt{sing}). We will probe the impact of not fixing LDCs in Sect. 4. For each light-curve fit, the residuals are computed and the maximum absolute residual (which should be at ingress or ingress) is calculated.

 Using the numerical tool SOAP2.0-T which does not account for the ring, we fitted each of the 63 simulated ringed planet RM signals with a ringless planet model.\footnote{Other analytical tools such as ARoME \citep{boue} could have been used for fitting the RM signals. However, this was not used because we are concerned only with the impact of the rings and do not want to be affected by the slight difference between the RM signals from ARoME and SOAP3.0. This difference arises from approximations used in the analytical tool \citep{boue,oshagh-2016}.} We started the fit initially with $\nu \sin i_{\ast}$, $\lambda$, $R_{p}$ and $i_{p}$ as free parameters while the other parameters were fixed to values in Table \ref{simul}. This was done to investigate if the presence of rings around an exoplanet could make a planet seem misaligned or cause an inaccurate estimation of $\nu \sin i_{\ast}$. The values of $\nu \sin i_{\ast}$ and $\lambda$ were found not to change and because performing a numerical fit is computationally intensive, we reduced the free parameters to $R_{p}$ and $i_{p}$. The residual of each RM signal fit was also computed and the maximum absolute residual calculated. 

In Fig. \ref{fits} we plot fits to the light-curves and RM signals of the simulated face-on and edge-on ringed planet transits. The plots also show the residuals with maximum amplitudes at ingress and egress. As expected, the edge-on ringed planet signals show no ring signature in the residuals. On the other hand, fits to the light-curve and RM signal of the face-on ring planet produces large residuals (3.15\,m/s for RM and 455\,ppm for flux) at ingress and egress with a duration of 70\,mins. The light-curve residuals show symmetry about mid transit phase (0.25) while the RM residuals show anti-symmetry. In order to accurately detect the ring signature, a photometric noise level  $\leq$100\,ppm and RV precision of 1\,m/s is required for each exposure. We therefore set 100\,ppm and 1\,m/s as the photometric and RV detection limits. These are reasonable limits based on the precisions of current and near-future instruments. These limits are shown in the residual plots. We explore the impact of these limits in Sect. 4.

As seen in Fig. \ref{fits}, the $R_{p}$ derived from the face-on ringed planet fit is greater than the actual planet radius used in Table \ref{simul} and this would lead to an underestimation of planetary density if planet mass were known.  Also, as shown in \citet{zuluaga} for the photometric transit, the $a$ and $i_{p}$ derived for most orientations of the ring will be underestimated (compare light-curve fit values in Fig. \ref{fits} to values in Table \ref{simul}). In turn, the transit-derived stellar density would be underestimated when compared to asteroseismology derived stellar density \citep{santos}. Perhaps, the discrepancies between both methods for derivation of stellar density \citep{huber} could be explained by the presence of planetary rings amongst other explanations \citep{kipping}. 

 The light-curve residuals of the edge-on ring produces some high-frequency irregularities around mid-transit phases with maximum absolute residual of 11\,ppm. This is due to numerical noise in SOAP3.0 light-curve computation (see also Fig. 3 of \citealt{oshagh-soapt}) but these irregularities are far below the detection limit considered. These mid-transit irregularities from the code are also present for the face-on ring and an additional non-linear trend is noticed in this region. This trend arises from the non-linear limb darkening law whose coefficients cannot compensate for the different inclination ($i_{p}$) derived from the fit \citep{barnes}. But since ring signatures are localised to ingress and egress phases, the trend around mid-transit phases (which is smaller than detection limit) does not interfere with the accuracy in measuring the ring signatures.
 
 \subsection{Identifying favourable ring orientations for detection}
It is important to identify the possible ring orientations that will favour detection. To do this, we perform the light-curve and RM fit for all 63 ring orientations taking note of the residuals in each case.

\begin{figure}[ht!]
        \centering
        \includegraphics[width=1\linewidth]{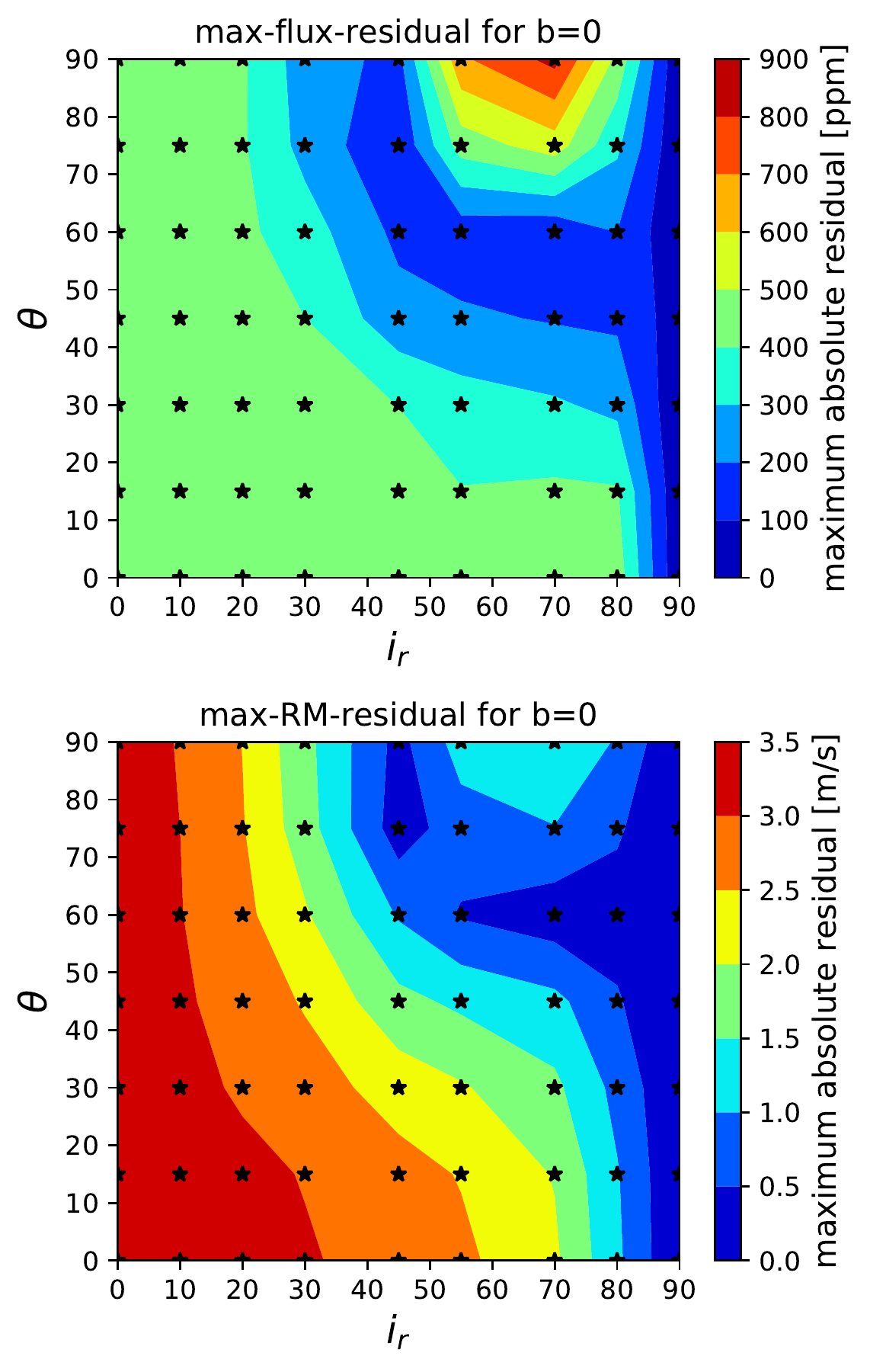}
        \caption{Contour plot from maximum absolute residual obtained from fit of 63 ring orientations. Top plot shows the contour plot for the light-curve fit while bottom plot shows the contour plot for the RM fit.}
        \label{contours}
\end{figure}

Figure \ref{contours} shows contour plots generated using the maximum absolute residuals of each ring orientation with the overplotted asterisks indicating the orientations from which the residuals were obtained. It is seen for the light-curve and RM residuals that several of the ring orientations favour easy detection.

For all $\theta$ values within $i_{r} \leq 30^\mathrm{o}$ (at and around face-on), the ring signatures are very prominent both for the light-curve and RM signal due to large stellar area covered by the ring. However, as $i_{r}$ increases up to $70^\mathrm{o}$ for the RM and $80^\mathrm{o}$ for light-curve, good detectability gradually shifts to only values of $\theta \leq 30^\mathrm{o}$. A blue band of low or undetectable ring signature for light-curve and RM signal is noticed for points inside about $i_{r} > 40^\mathrm{o}$ and $\theta >40^\mathrm{o}$ up to all edge-on ($i_{r} \simeq 90^\mathrm{o}$) orientations. This indicates that transit signal with ring orientation within this blue region can nearly be approximated by a ringless transit model.

A separate region with high ring signature is evident in the residual of the light-curve fit centred around $i_{r},\theta=70^\mathrm{o},90^\mathrm{o}$. Around this orientation, the high $i_{r}$ causes only a small projected ring area whereas $\theta=90^\mathrm{o}$ makes the ring perpendicular to orbital plane. This causes the transit duration to be the same as that of a ringless planet but there would be an increased transit depth. A ringless fit is unable to perfectly reconcile the transit duration with the depth thereby causing an overestimation of the duration which leads to a high residual at ingress or egress. This is also observed for the RM residual in this region although not nearly as significant as that of the light-curve. The plot of this orientation is shown in Fig. \ref{70,90}.\\

It is therefore seen that a lot of the ring orientations favour detection although most of them are close to face-on where a greater stellar area is covered by the ring. However, for close-in ringed planets, the inclination $i_{r}$ is expected to have been damped towards edge-on due to tidal forces from the star making ring detection difficult \citep{sch11,heising}. \citet{sch11} also showed that exoplanets with $a > 0.1$ AU can have $i_{r}$ values that favour detection. If we take $i_{r} \geq 45^\mathrm{o}$ to represent feasible exoplanetary ring inclinations for this close-in planet then we have quite a number of orientations in Fig. \ref{contours} with high ring signatures interesting for the search for exoplanetary rings. We will explore how the ring signatures change with impact parameter in Sect. 4.

\begin{figure}[t]
        \centering
        \includegraphics[width=1\linewidth]{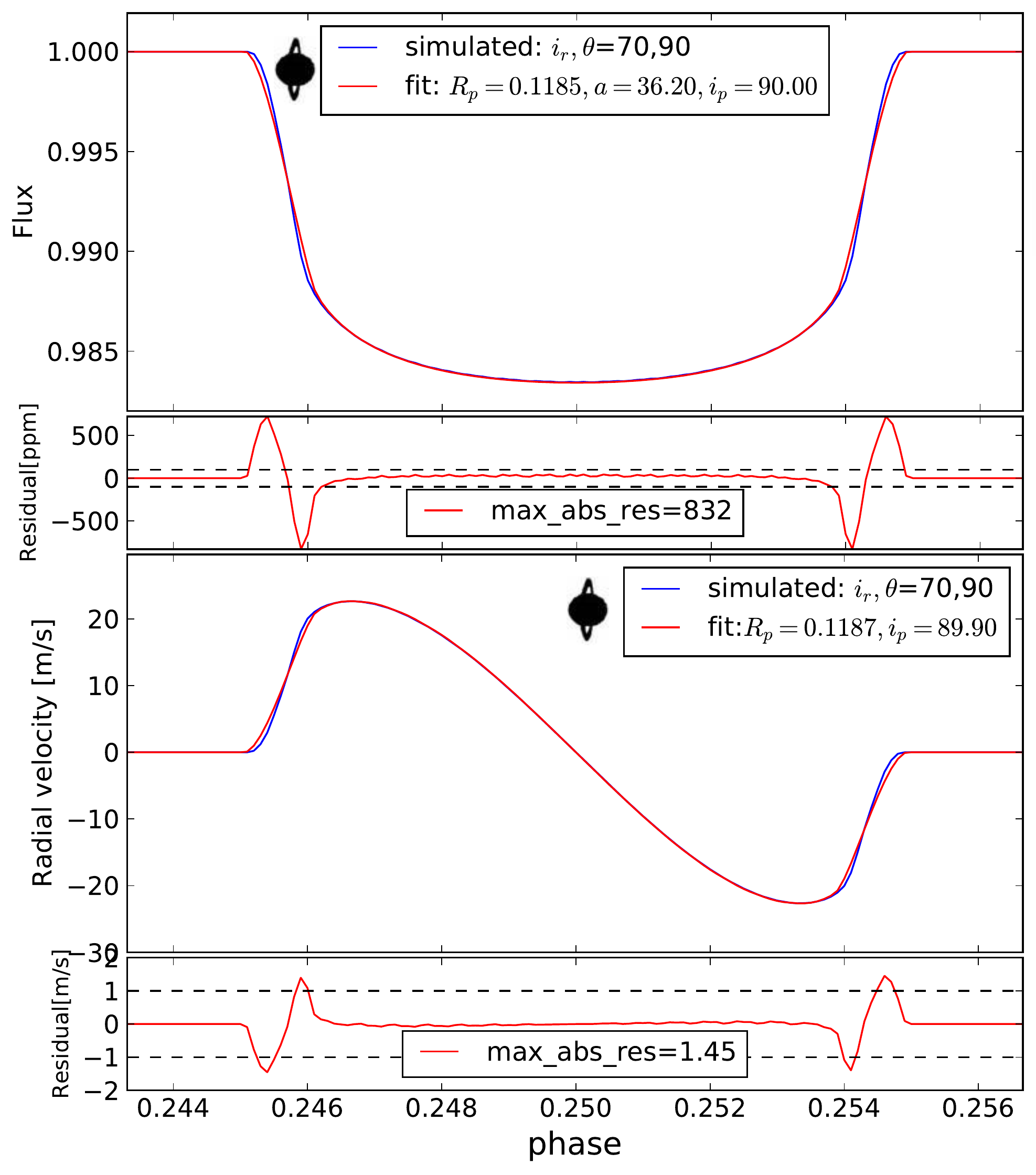}
        \caption{Light curve and RM signal fit for $i_{r},\theta=70^\mathrm{o},90^\mathrm{o}$.}
        \label{70,90}
\end{figure}

\section{Discussion}
In this section we describe the impact of some assumptions and other effects that may come into play in the detection of exoplanetary rings.
\subsection{Effect of ring-planet gap and ring area}

\begin{figure*}[!ht]
\centering
\begin{subfigure}{.5\textwidth}
  \centering
  \includegraphics[width=1\linewidth]{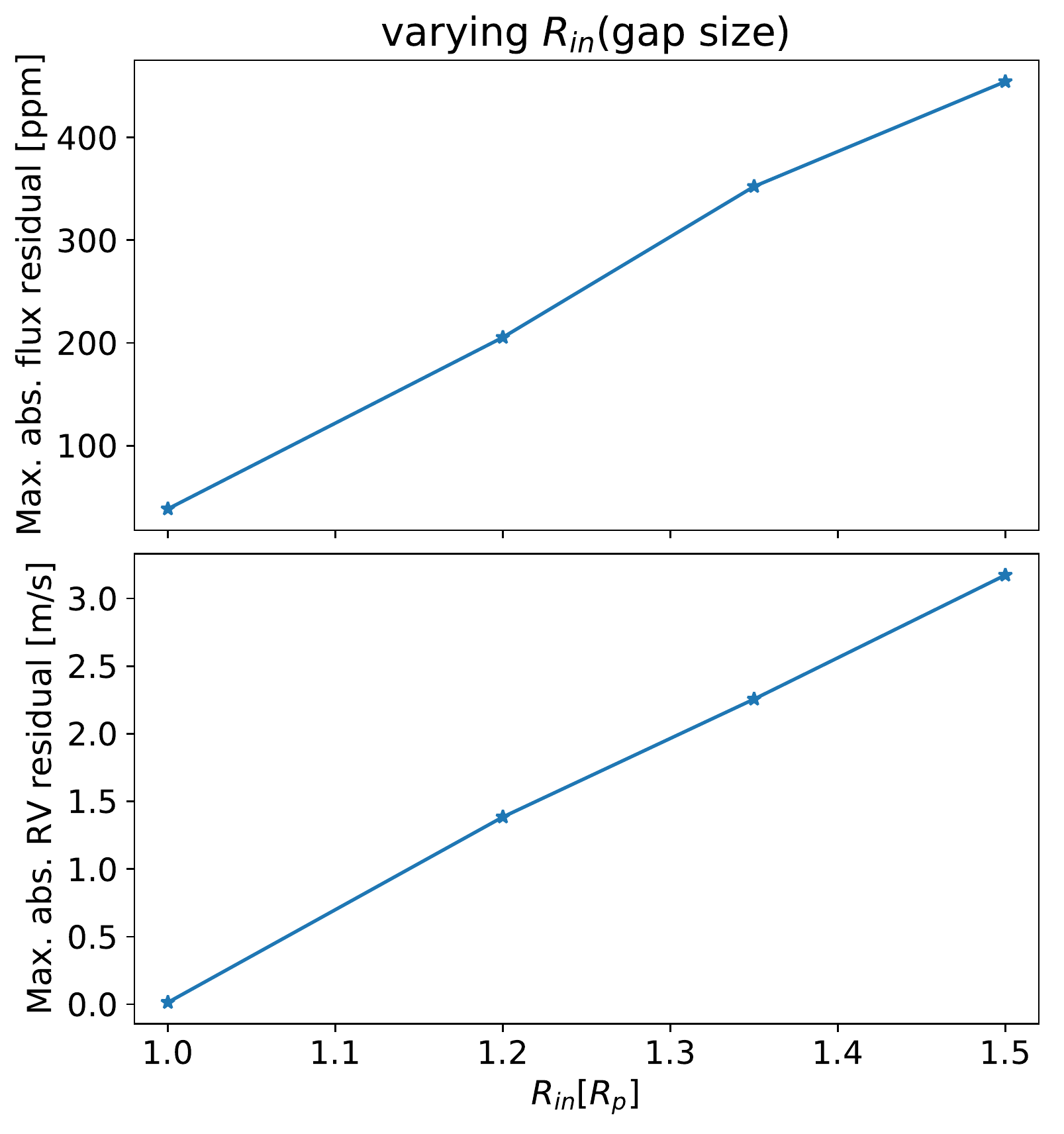}
\end{subfigure}%
\begin{subfigure}{.5\textwidth}
  \centering
  \includegraphics[width=1\linewidth]{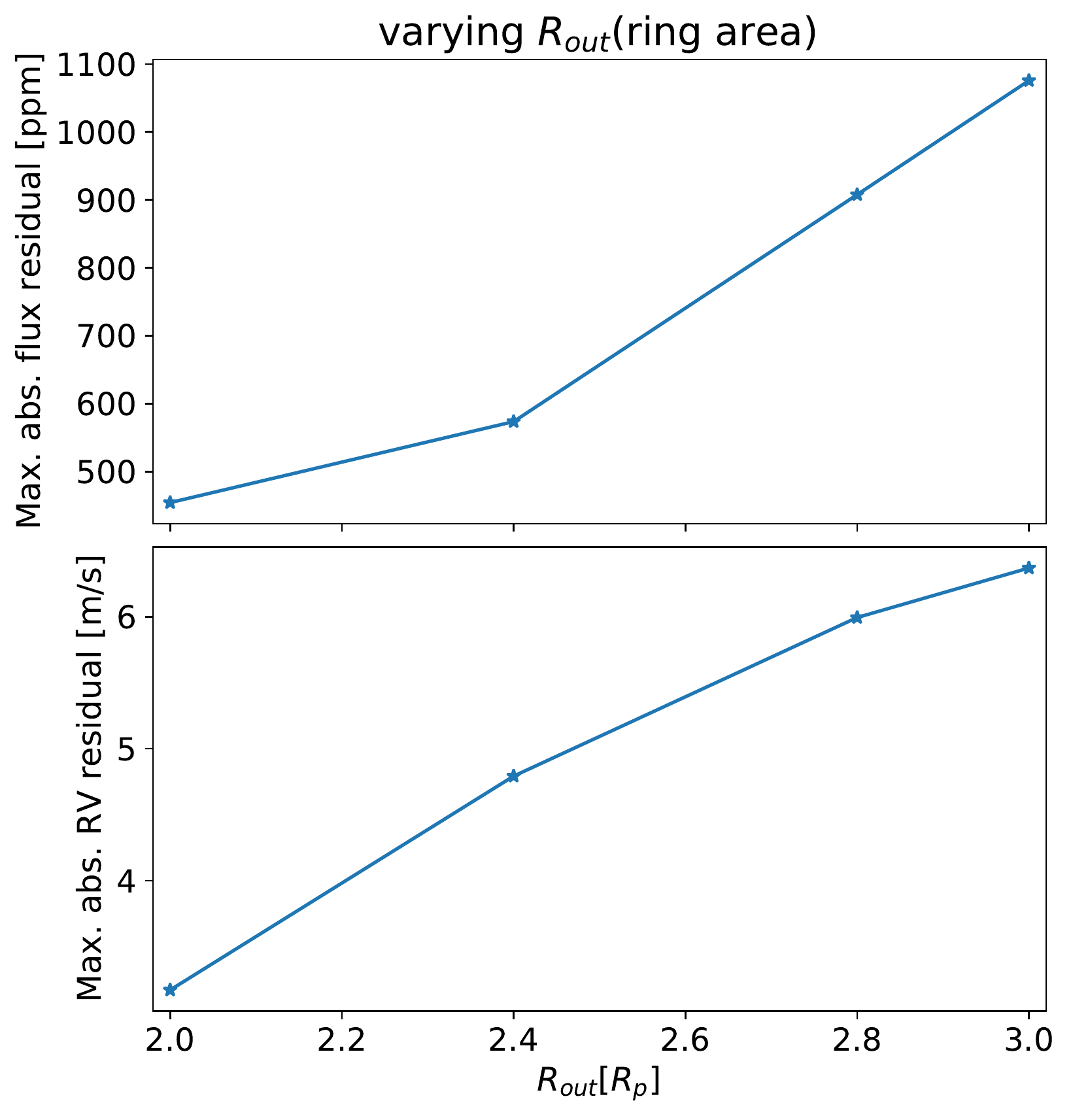}
\end{subfigure}
        \caption{Left column: Effect of planet-ring gap on ring signature by increasing $R_{in}$. Right column: Effect of ring area on ring signature by increasing $R_{out}$. Top plots are the effects in flux and bottom plots are the effects in RV.}
\label{ring_area_and_gap}
\end{figure*}

As seen in the third column of Fig. \ref{varying}, the gap between ring and the planet's surface and also the variation of the ring area changes the ringed planet's signal. We assess here how these changes impact the ring signature. To assess the gap impact we selected the face-on ring orientation and vary the value of $R_{in}$ from $1\,R_{p}$ up to $1.5\,R_{p}$ while maintaining a constant ring area.  The left column of Fig. \ref{ring_area_and_gap} shows that as the planet-ring gap increases, the ring signature also increases both for the flux and RM. When ring is in contact with planet surface at face-on, the ringed planet signal is the same as that of a ringless planet with a larger radius since the ring here is optically thick. However, if the ring is only nearly face-on, the transit light-curve would be identical to that of a very oblate planet \citep{barnes03}.

To assess the impact of ring area, we kept $R_{in}=1.5\,R_{p}$ while increasing the value of $R_{out}$ from $2\,R_{p}$ up to $\,3R_{p}$. The right column of Fig. \ref{ring_area_and_gap} shows that the ring signature also increases with ring area in the flux and RM. As ring area increases, the transit signal of the ringed planet gets increasingly similar to the grazing eclipse of a binary star having a V-shaped light-curve.
Thus our finding suggests that it is easier to detect rings with larger planet-ring gap and larger ring area.

\subsection{Effect of limb darkening}

\begin{figure*}[th!]
        \centering
        \includegraphics[width=1\linewidth]{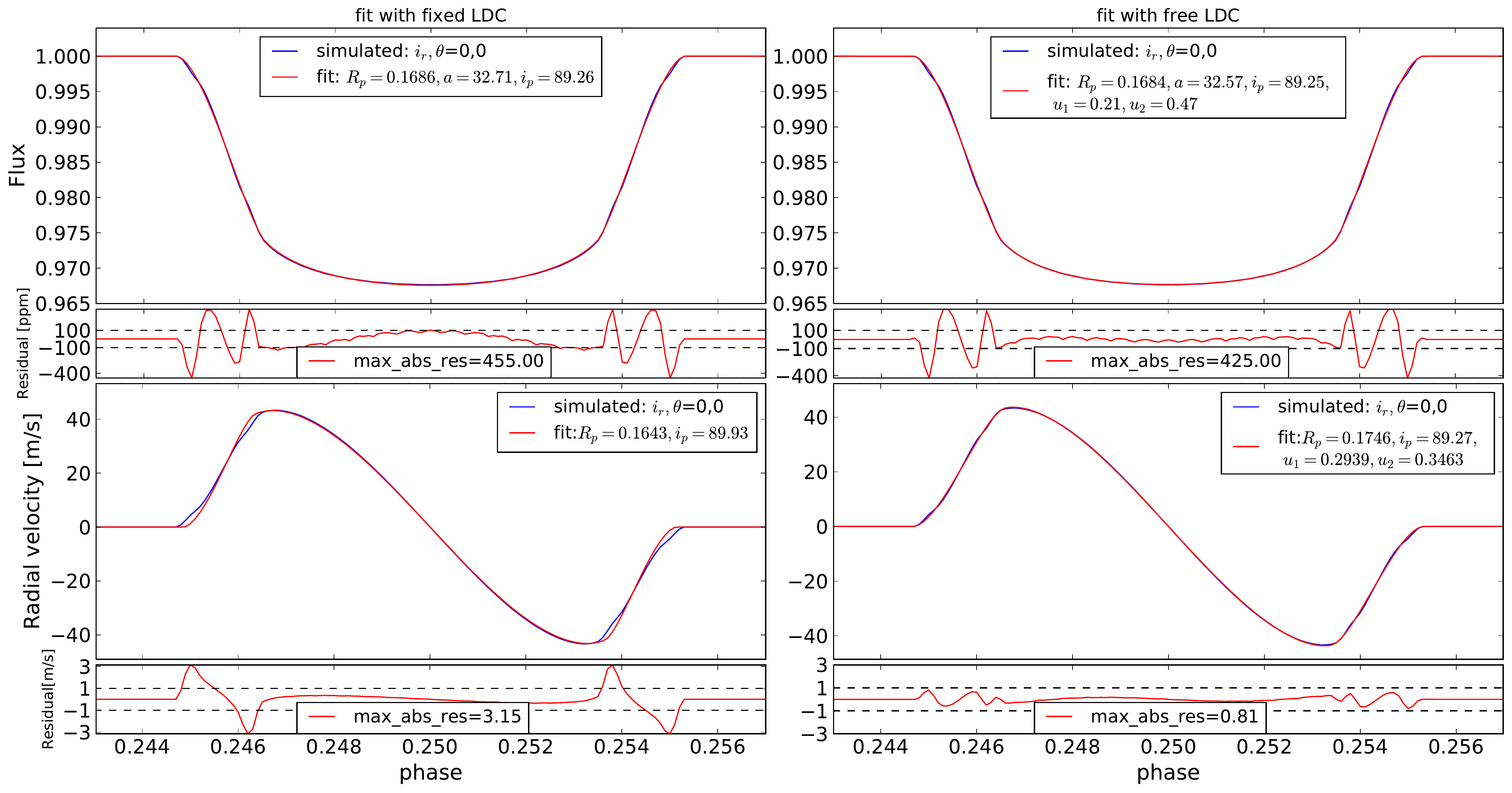}
        \caption{Effect of fitting ringed planet signal with constant LDC and free LDC. Left column: Fit of face-on ring with LDC kept constant at $u_{1}$=0.29, $u_{2}$=0.34 for light-curve and RM signal fit. Right column: Fit with LDC allowed to vary.}
        \label{ldcfit}
\end{figure*}

For most fitting procedures, if the limb darkening coefficients (LDCs) are known a priori they are fixed during the fitting as was done in Sect. 3. This reduces the number of free parameters thereby increasing accuracy of the results and could eliminate some degeneracy in the fitting process. However, limb darkening affects transit signals at ingress, egress and signal amplitude and so can compete with the ring signature. Therefore, we assessed impact of inaccurate estimation of LDCs ($u_{1}$,$u_{2}$) on the detection of the ring signature both in the light-curve and RM signal by fitting them as free parameters. We used the face-on ring orientation for this test and the result is shown in Fig. \ref{ldcfit}.\\ 

It is seen that fitting the LDCs give rise to an inaccurate estimation of the LDCs. Comparing the free-LDC fit residual to the fixed-LDC fit residual shows that the inaccurate estimation leads to damping of the ring signature at ingress and egress. For the light-curve fit, the estimated LDCs are different from the values used for the simulated ringed planet but the residual ring signature is damped only by a small amount. For the RM signal fit, the estimated LDCs are close to that of the ringed planet yet there is a significant damping of the ring signature below the detectable limit of $1\,m/s$. Therefore, inaccurately estimating limb darkening parameters has greater effect in RV and can render ring signatures undetectable. 

It has been shown that fitting LDCs in transit analysis can lead to LDCs different from the theoretical LDC calculated from stellar evolution models (see \citealt{barros,neil}). Therefore, for very high precision transits a very careful modelling of the LDCs needs to be performed so that the LDCs don’t bias the results \citep{csiz}. As discussed in \citet{pav15}, LDCs can be constrained from informative priors based on modern tabulated LDCs gotten from advanced spherical stellar atmosphere models (e.g. from \citealt{sing,claret14,husser}). Using the Limb darkening toolkit by \citet{pav15}, the typical LDC uncertainties are $\sim$0.2 - 0.5\% and are obtained from propagated uncertainties in the estimation of the stellar parameters. When the LDCs are constrained with these uncertainty values, detectable ring signature is again recovered. This method provides a compromise  between fixed LDCs and totally unconstrained LDCs (which as seen in Fig \ref{ldcfit} can mask the ring signature).

\subsection{Effect of stellar rotation velocity}
\begin{figure}[t!]
\centering
\includegraphics[width=1\linewidth]{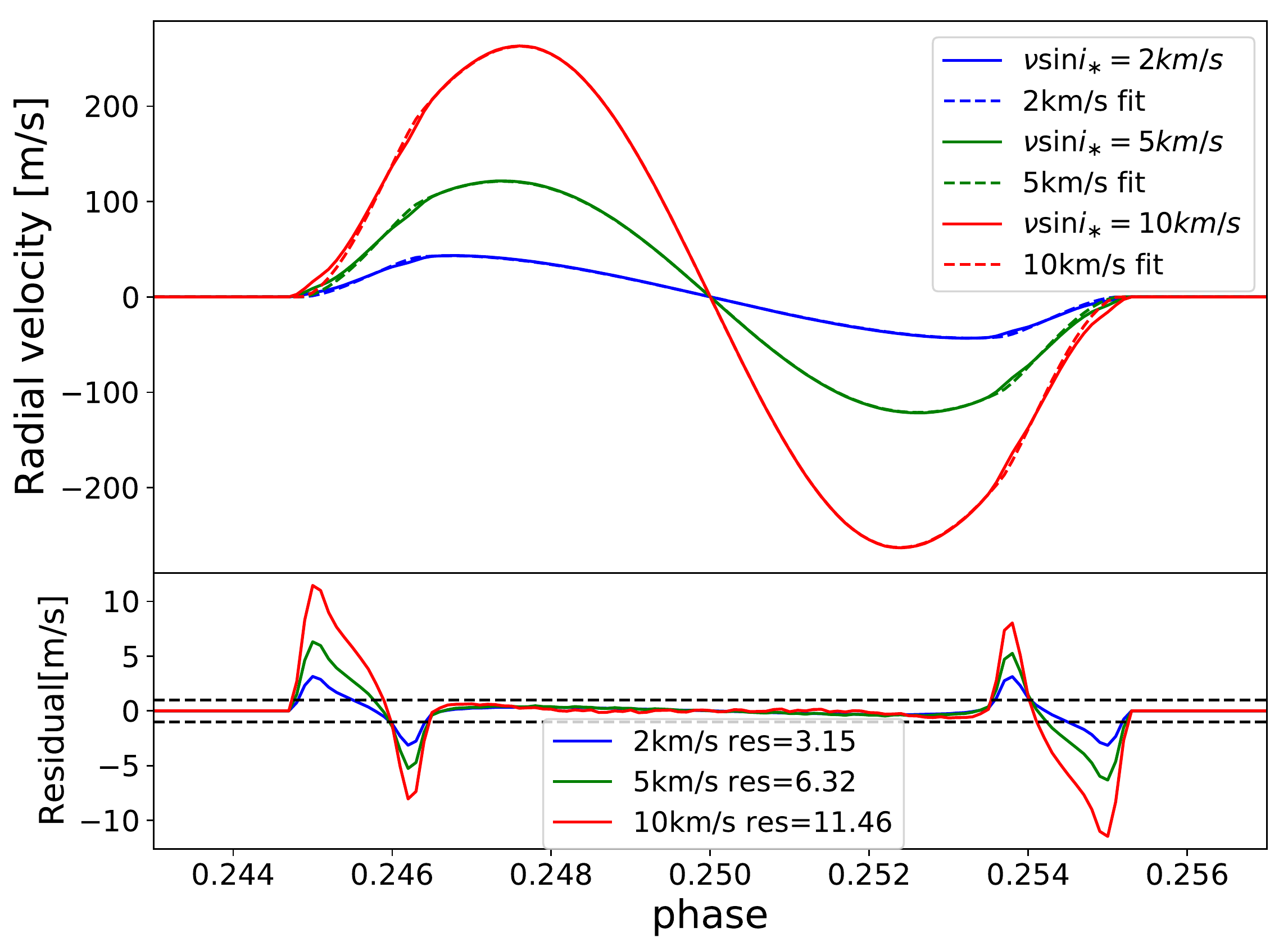}
\caption{RM signal for different stellar rotation velocity fitted with a ringless model and their computed residuals in m/s}
\label{vary_vsini}
\end{figure}

As the RM effect is proportional to the stellar rotational velocity $\nu \sin i_{\ast}$, we needed to assess how this impacts the RM ring signature. We again used the face-on ring orientation with stellar rotations of 2, 5 and 10\,km/s and perform fits to the generated ringed signals to calculate the residuals. Figure \ref{vary_vsini} shows that not only the RM signals increase with $\nu \sin i_{\ast}$ but also the ring signatures and so it can be easier to detect rings around planets transiting fast rotating stars. However, a fast stellar rotation velocity causes broadening of spectral lines which will degrade the RV precision so a compromise has to be made between stellar rotation velocity and needed RV precision.

\subsection{Effect of planet orbital inclination (impact parameter)}

\begin{figure}[ht!]
\centering
\includegraphics[width=1\linewidth]{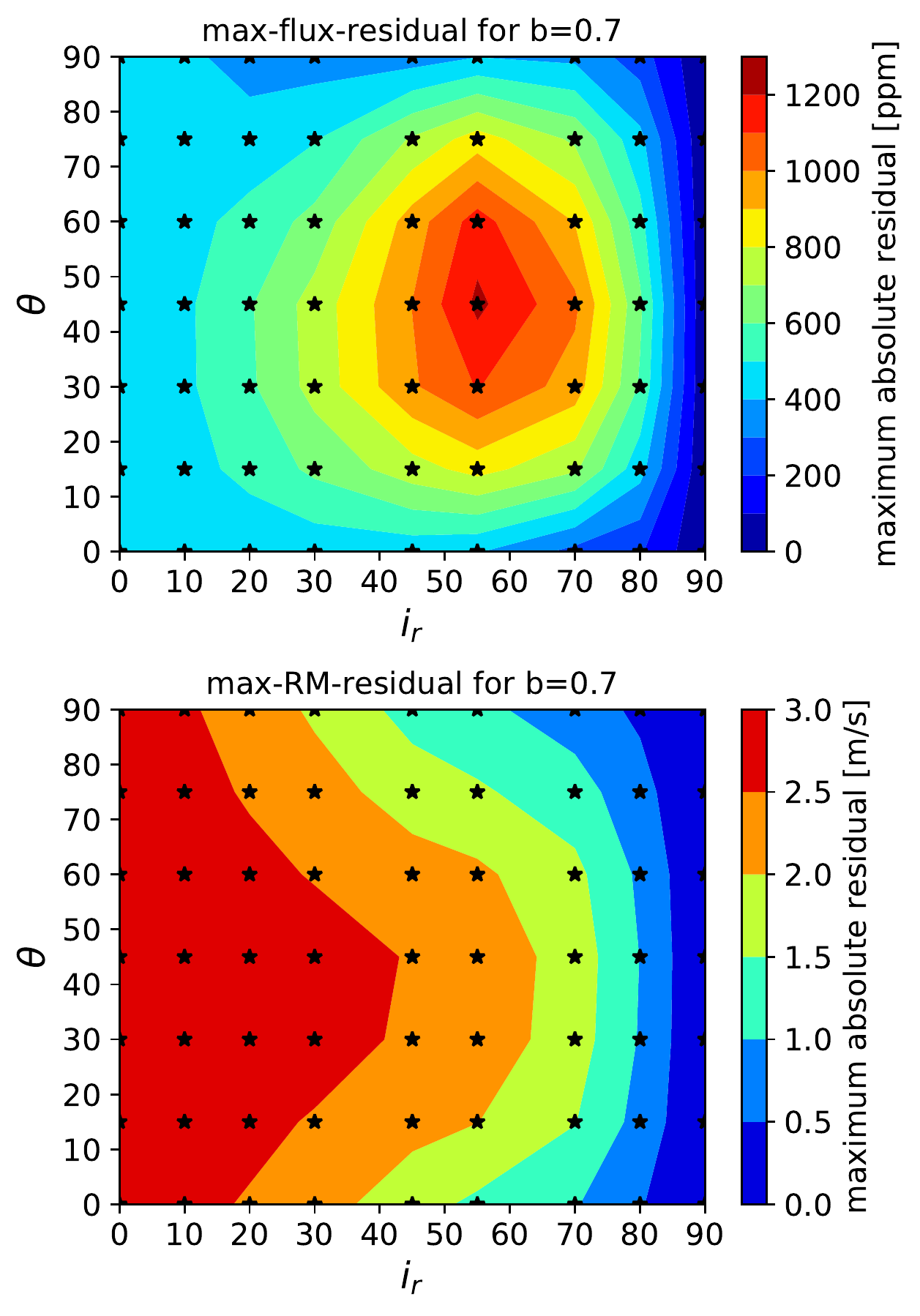}
\caption{Contour plot from maximum absolute residual obtained from fit of 63 ring orientations at $b=0.7$. Top plot shows the contour plot for the light-curve fit while bottom plot shows the contour plot for the RV fit}
\label{b,07}
\end{figure}

\begin{figure}[h!]
\centering
\includegraphics[width=1\linewidth]{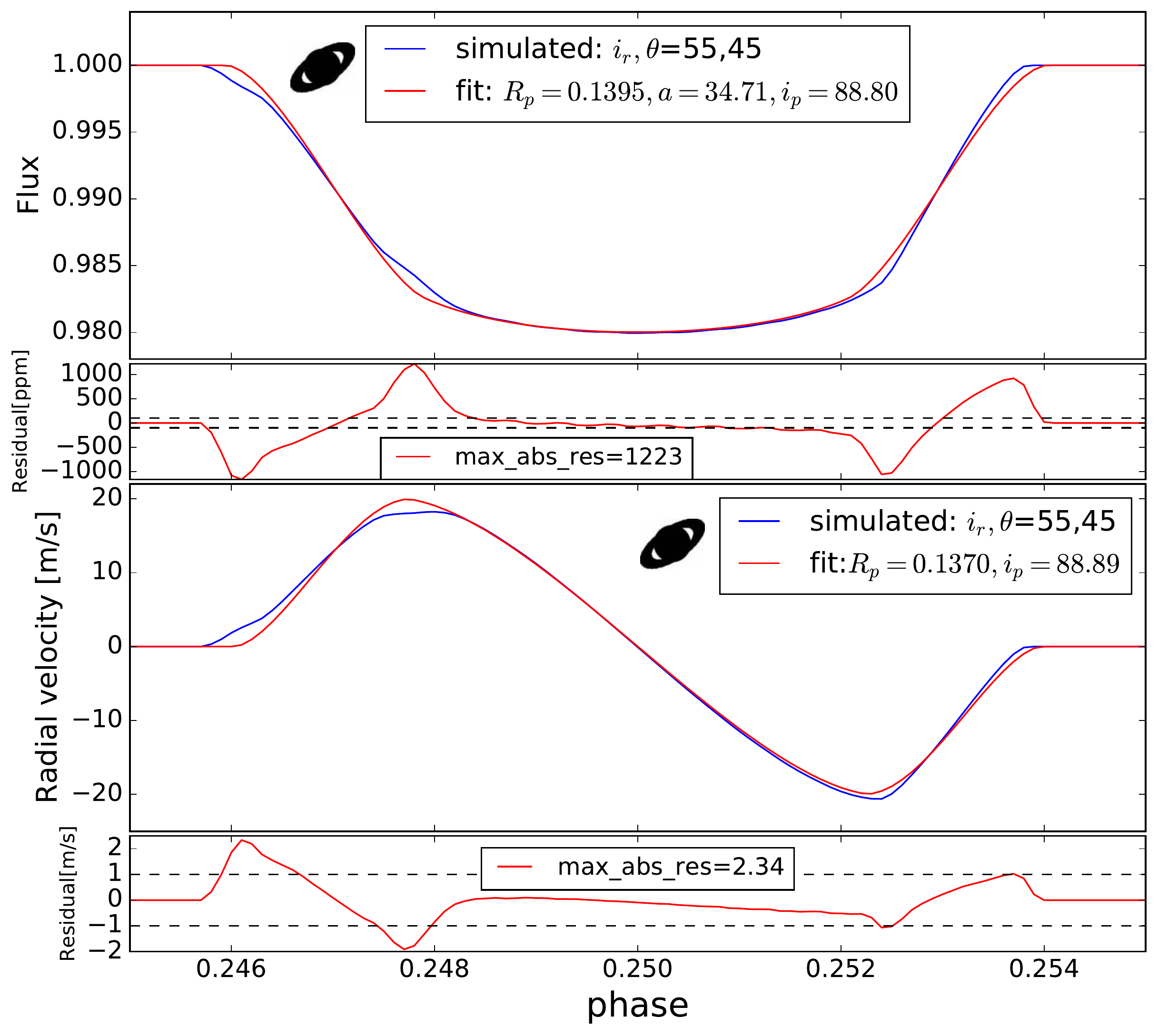}
\caption{Asymmetric light-curve and RM signal of $i_{r},\theta=55,45$, the ringless fit and residual}
\label{5545}
\end{figure}

The results in Sect. 3 were obtained for a planet with orbital inclination $i_{p}$=$90^\mathrm{o}$ translating to impact parameter $b$=0. Here we investigate the ring signatures at impact parameter of 0.7 ($i_{p}$=$88.89^\mathrm{o}$) by making same contour plot for $b$=0.7 as was done for $b$=0 in Fig. \ref{contours}. The contour plot for $b$=0.7 is shown in Fig. \ref{b,07}. It is seen that ring signatures are high even so close to edge-on at $i_{r}=80^\mathrm{o}$ for flux and $i_{r}=70^\mathrm{o}$ for RM. It is seen in the flux residual from light-curve fit that ring signature is highest at $i_{r},\theta=55^\mathrm{o},45^\mathrm{o}$ and reduces radially from that orientation but goes to zero at edge-on orientations. The high residual at $i_{r},\theta=55^\mathrm{o},45^\mathrm{o}$ is due to asymmetry in the light-curve caused by the ring tilt, high impact parameter, and stellar limb darkening. The light curve is asymmetric because the upper (leading) part of the ring blocks brighter stellar regions during ingress while the lower (trailing) part of the ring blocks a relatively darker region as the ringed planet exits the stellar disc (an illustration of high $b$ transit can be seen in \citealt{heising}). This asymmetry is responsible for high residuals around points where $\theta \neq 0$ or 90.

 The residual from RM signal is highest at and around face-on orientations and reduces gradually to zero towards edge-on orientations. There is also asymmetry in the RM signal for the same reason as above (different parts of the ring blocking stellar regions with different RV components). The asymmetry is prominent especially around $\theta=45^\mathrm{o}$ causing a higher residual at $30 \leq \theta \leq 60$ than other $\theta$ at same $i_{r}$. The light-curve and RM signal for the $i_{r},\theta=55^\mathrm{o},45^\mathrm{o}$ orientation is plotted in Fig. \ref{5545}.

\subsection{Impact of time-sampling and instrument precision}

For our fiducial planet (Table \ref{simul}) on 25-day orbit (0.16\,AU), the maximum transit duration (with face-on ring) is 6.38\,hrs and the duration of the ring signature at ingress or egress is $\sim$\,70\,mins. The detection of the ring signature in this timescale will require a high precision and high time resolution.

\begin{figure}[t]
        \centering
        \includegraphics[width=1\linewidth]{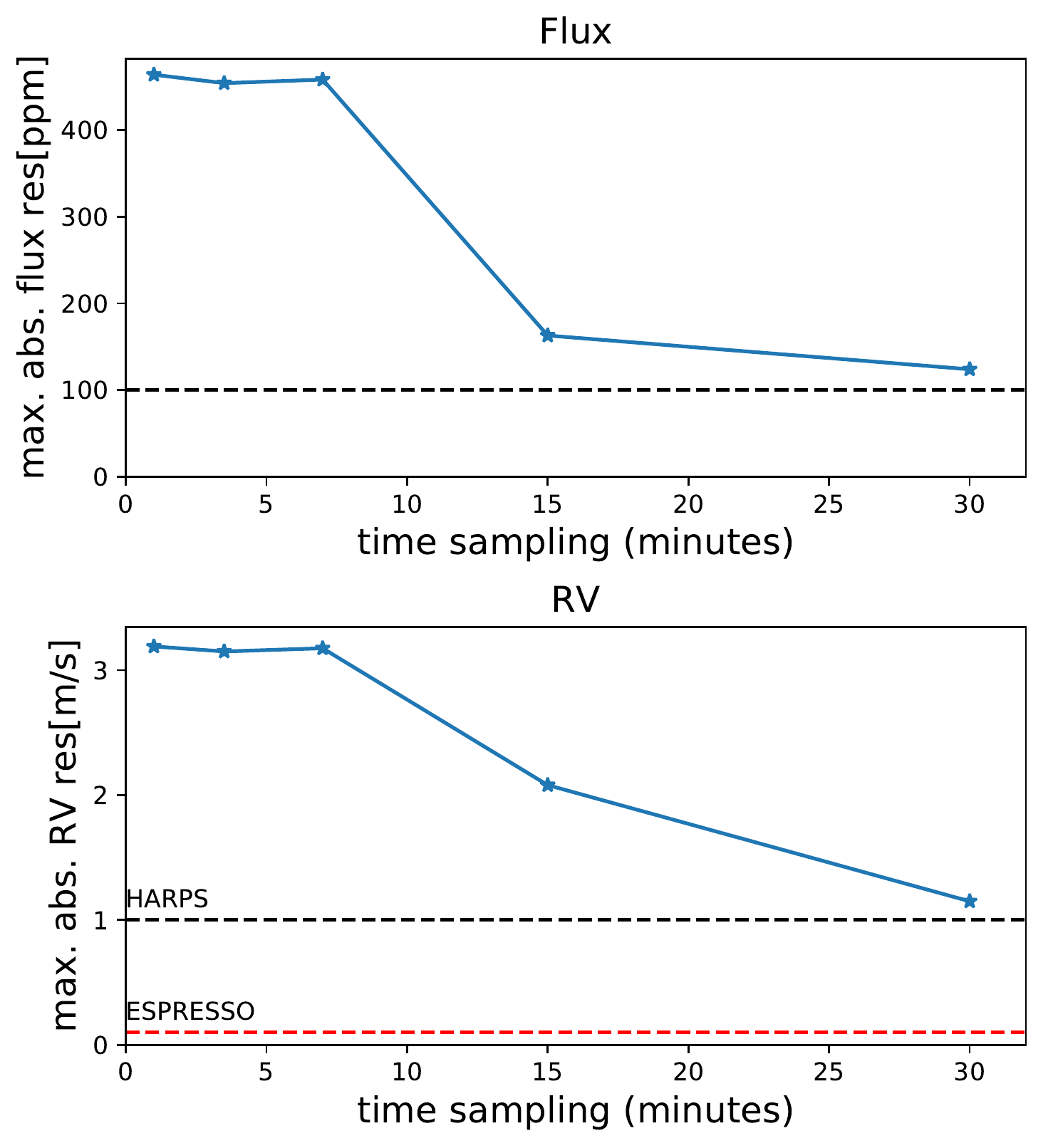}
        \caption{Top pane: Amplitude of photometric ring signature for different time-sampling. Black line indicates the detection limit of 100\,ppm. Bottom pane: Amplitude of spectroscopic ring signature for different time-sampling. Black and Red dashed line indicate detection limit of HARPS (1\,m/s) and ESPRESSO (0.1\,m/s).}
        \label{time_resplot}
\end{figure}

Here we assess the impact of time-sampling and instrument precision on ring signature detection. For long time-sampling, the ring signature might not be well-captured (under-sampled) with the few observational data points whereas for short time-sampling, the noise level per point might be too high to detect the ring signature. Therefore, a compromise has to be reached between the time-sampling and the achieved photon noise limited precision especially in the photometry. Using our fiducial planet with face-on ring, we simulate transit signals with different time-sampling from 30\,mins long-cadence to 1\,min short-cadence (30, 15, 7, 3.5 and 1 min). The fitting procedure done in Sect. 3 is again performed on each of these transit signals and we generate the residuals. Figure \ref{time_resplot} shows the maximum absolute residual obtained for the different time-sampled signals. For each of the time-sampled signals, the precision attained for each exposure should be at/below the set detection limit.

\begin{figure}[t]
        \centering
        \includegraphics[width=1\linewidth]{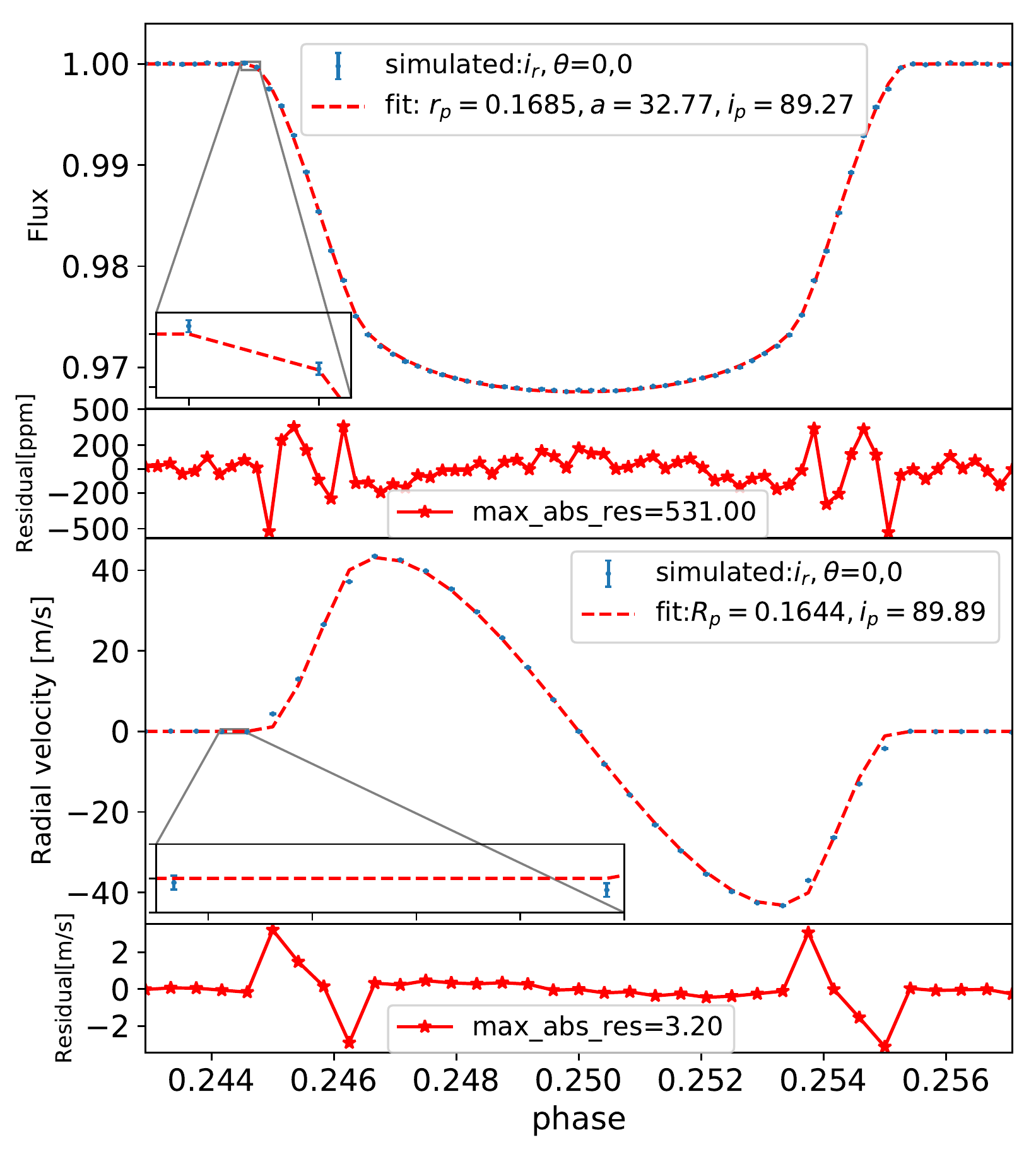}
        \caption{Top pane: Fit of simulated light curve with \textit{CHEOPS} noise level of 56\,ppm for time-sampling of 7\,mins. Bottom pane: Fit of simulated RM signal with noise at \textit{ESPRESSO} level of 0.1\,m/s and 15\,mins time-sampling. Insets show the zoom of the error bars.}
        \label{error}
\end{figure}

It is seen in the flux residual plot that the ring signature is prominent (above the 100\,ppm detection limit) for time-sampling below 15 minutes. This confirms the photometric result from \citet{barnes} that the detection of large rings require $\sim$15\,min time-sampling. However, the best photometric results are gotten with time-sampling between 1 - 7 mins where the 70\,min ring signature is well sampled. This is within the time-resolution of the upcoming instrument \textit{CHEOPS}\footnote{CHaracterising ExOPlanet Satellite. \textit{CHEOPS} will have photon noise limited precision of 150 ppm/min for transit across a G5 dwarf star of V=9 magnitude \citep{broeg}.}.
The ring signature is fairly constant below 7\,mins time-sampling indicating that 7\,mins suffices for the ring detection. The precision of \textit{CHEOPS} in 7\,mins is 56\,ppm (for 9th magnitude star) which will allow the detection of even low amplitude ring signatures. A time-sampling of 3.5\,mins will be required for orientations where the ring signature has a short duration. The expected light curve for face-on ringed planet is simulated in top panel of Fig. \ref{error} with 7\,mins time-sampling. The error bars were obtained by adding random Gaussian noise at the level of 56\,ppm (\textit{CHEOPS} precision in 7\,mins). The residual of the fit to the simulated light curve shows prominent ring signature at ingress and egress. 

At 15 minute time-sampling, the RV ring signature is well above the 1\,m/s detection limit. We note that 1\,m/s is the current RV precision of \textit{HARPS}\footnote{High Accuracy Radial velocity Planet Searcher} \citep{mayor}. This is very promising for the spectroscopic search for rings since RV measurements typically require up to $\sim$15\,min integration to average out the short-period stellar oscillations on FGK stars. In addition, upcoming spectrographs like \textit{HIRES} on E-ELT \citep{marconi} and \textit{ESPRESSO}\footnote{Echelle SPectrograph for RockyExoplanets and Stable Spectroscopic Observations} on VLT will present interesting possibilities for ring detection. For instance, \textit{ESPRESSO} will be capable of 0.1\,m/s accuracy on fifth magnitude stars in 1\,min exposures \citep{pepe}. Although the typical $\sim$15\,min RV exposure might preclude a 1\,min time-sampling, the unprecedented accuracy will increase ring detectability. The \textit{ESPRESSO} detection limit of 0.1\,m/s is also shown in the bottom plot of Fig. \ref{time_resplot}. However, it is not clear if this sort of RV precision can be achieved for G and K stars due to stellar granulation and oscillation noise \citep{dumus11}. The expected RM signal is simulated in bottom pane of Fig. \ref{error} with 15\,mins time-sampling. The error bars are obtained by adding random Gaussian noise at the \textit{ESPRESSO} level of 0.1\,m/s. The residual of the fit again shows ring signature at ingress and egress.

The duration of the ring signature depends on semi-major axis, so for the same ringed planet but with a semi-major axis of $a_{pr}$, the duration of ring signature will be $70 \times (a_{pr}/0.16$\,AU)$^{1/2}$. This implies that it will be easier to detect rings around longer-period planets since the ring signature timescale will be longer and can be sampled more easily. For instance, our ringed planet at 1\,AU will have ring signature timescale of 175\,mins. The resulting light-curve can still be well-sampled with 15\,min exposure time and a photometric precision of 39\,ppm (for ninth magnitude star) will be achieved with \textit{CHEOPS} while the RM signal can be sampled with up to 25\,min exposure. The impact parameter, planet radius and ring parameters can also alter the duration of the ring signature.

\section{Conclusions}

This paper has introduced SOAP3.0, a numerical tool for the simulation of photometric and spectroscopic signal of a transiting ringed planet. It is capable of generating light-curves and RM signals with the additional effects of rings included. We confirmed suitability of this tool by comparing its result with others in literature. We used the tool to characterise ring signatures considering different possible orientations of the ring and we showed the ring orientations that are favourable for detection using each transit technique. Most interesting are the orientations close to edge-on for which ring signatures can still be detected. This is very promising in the search for exoplanetary rings and the characterisation sheds important light on how rings might be detected around exoplanets if they exist as expected. 

We have shown different factors that would impact both methods either to amplify or attenuate the ring signature. We observe that: 
\begin{itemize}
    \item The gap between planet surface and ring inner radius is pertinent for ring detection.
    \item High impact parameter transits cause asymmetry in the signals which leads to large ring signatures for tilted rings.
    \item Transits across fast rotating stars can have more prominent spectroscopic ring signatures than in the photometry.
    \item Inaccurate estimation of the limb darkening coefficients leads to damping of ring signatures more in the spectroscopic RM signals than in the photometric light-curves.
    \item Time-sampling $\leq$7 minutes is required for the photometric ring detection while 15\,minute sampling suffices for spectroscopic ring detection.
    \item The precision of upcoming observational instruments like \textit{CHEOPS} and \textit{ESPRESSO} will increase the ring detectability.
    
\end{itemize}

\noindent We restate that although we considered a planet orbiting at 0.16 AU, the method is valid for planets at any distance from the star with the only difference being in the timescale. Our results have thus shown the complementarity of the two transit techniques, a synergy which will increase the certainty of any positive ring detection. 
\begin{acknowledgements}
This work was supported by Fundação para a Ciência e a Tecnologia (FCT, Portugal) through national funds and by FEDER through COMPETE2020 by these grants UID/FIS/04434/2013 \& POCI-01-0145-FEDER-007672 and PTDC/FIS-AST/1526/2014 \& POCI-01-0145-FEDER-016886. NCS and SCCB also acknowledge support from FCT through Investigador FCT contracts IF/00169/2012/CP0150/CT0002 and IF/01312/2014/CP1215/CT0004 respectively. BA acknowledges support from  Centro de Investigação em
Astronomia/Astrofísica da Universidade do Porto (CAUP) in form of research fellowship contract with reference CIAAUP-08/2017-BI. MO acknowledges research funding from the Deutsche Forschungsgemeinschft (DFG, German Research Foundation)-OS 508/1-1 and also acknowledges the support of COST Action TD1308 through STSM grant with reference Number:STSM-TD1308-050217-081659.
\end{acknowledgements}

\bibliographystyle{aa} 
\balance
\bibliography{references} 

\onecolumn
\begin{appendix}

\section{Occultation of stellar spot by a ringed planet}

\begin{figure*}[th!]
        \centering
        \includegraphics[width=0.96\linewidth]{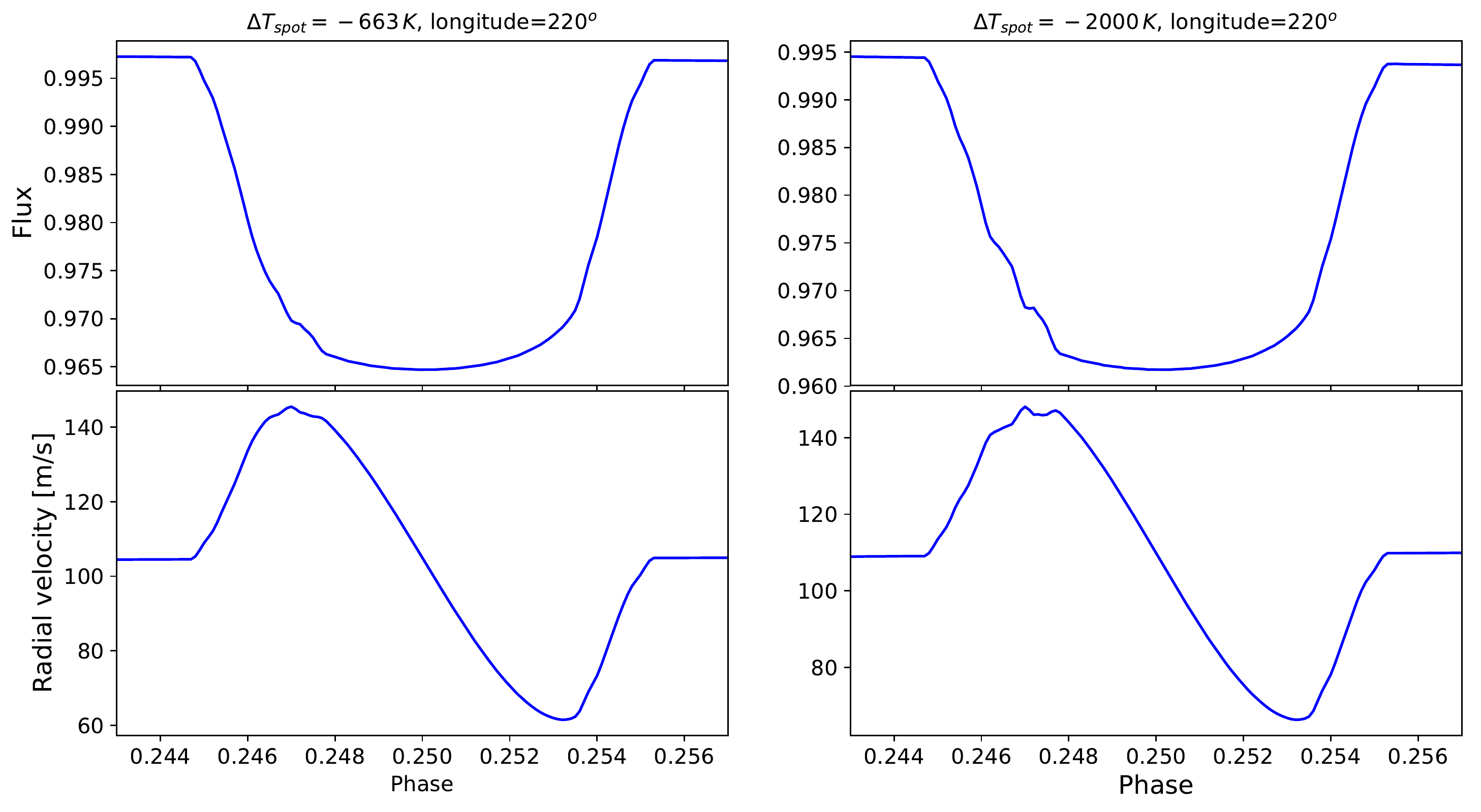}
        \caption{Ringed planet occultation of a stellar spot of 1\% filling factor ([$R_{spot}/R_{\ast}]^2$) during transit. The spot is positioned at latitude $0^o$ and longitude $220^o$ such that the occultation occurs within the ingress phase. Left panes show the light curve and RM signal when contrast ($\Delta T_{spot}$)  between spot temperature and star's effective temperature is -663\,K. Right panes are the signals when $\Delta T_{spot}=-2000\,K.$}
        \label{long220}
\end{figure*}

\begin{figure*}[h!]
        \centering
        \includegraphics[width=0.96\linewidth]{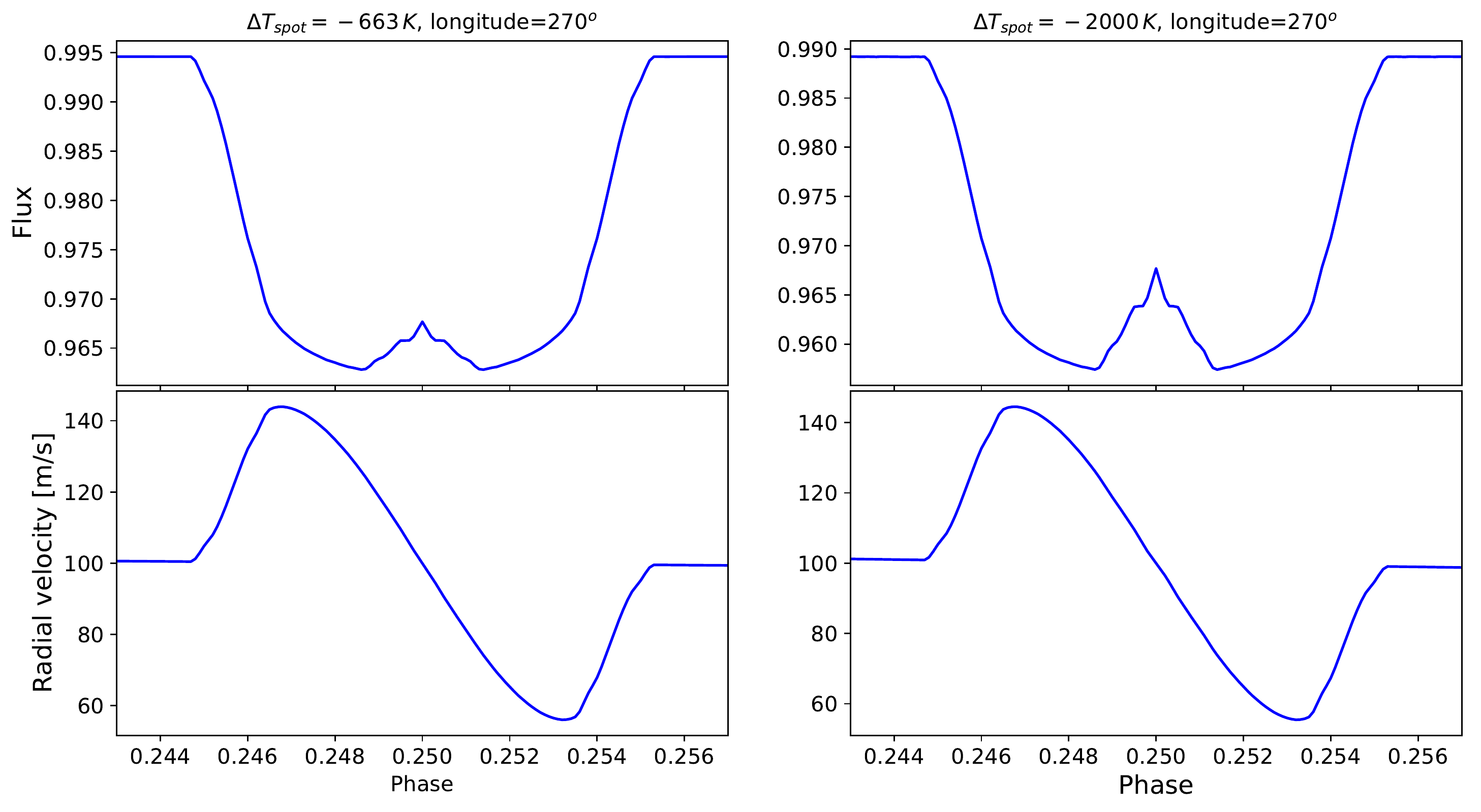}
        \caption{As Fig. \ref{long220} but spot longitude is $270^o$ such that the occultation occurs at midtransit. The effect of spot occultation is not visually prominent in the RM signal since RV values at stellar centre is essentially same as the out-of-transit value (stellar centre has zero radial velocity).}
        \label{long270}
\end{figure*}

\section{Table of SOAP3.0 comparison with other tools}
\begin{table*}[h!]
\centering
\caption{Comparison of SOAP3.0 with \textit{EXORINGS} using Table \ref{simul} input. Also comparison with quoted values of \citet{tsunski} using input values from the paper. SOAP3.0(LD) corresponds to results when limb darkening is used. Asterisk ($^*$) denotes orientations where the transit duration of \textit{EXORINGS} and SOAP3.0 differs.}  
\label{comparetable}
\addtolength{\tabcolsep}{7pt}
\begin{tabular}{c c c c c c c }
\hline\hline
\multicolumn{2}{c|}{Ring angles}   & \multicolumn{3}{c|}{Transit depth {(}ppm{)}}   & \multicolumn{2}{c}{Total transit duration {(}hours{)}} \\ \hline
$i_{r}$          & $\theta$         & \textit{EXORINGS}      & SOAP3.0      & SOAP3.0(LD)     & EXORINGS                    & SOAP3.0                   \\ \hline
0                & 0                & 27500         & 27515        & 32387           & 6.380                       & 6.383                     \\ 
20               & 0                & 26444         & 26450        & 31144           & 6.380                       & 6.383                     \\ 
45               & 0                & 22374         & 22387        & 26364           & 6.380                       & 6.383                     \\ 
70               & 0                & 14047         & 14052        & 16567           & 6.380                       & 6.383                     \\ 
$^{\ast}$90               & 0                & 10000         & 10000        & 11792           & 6.380                       & 5.850                     \\ 
$^{\ast}$0                & 45               & 27500         & 27515        & 32387           & 6.026                       & 6.383                     \\ 
$^{\ast}$90               & 45               & 10000         & 10000        & 11844           & 6.024                       & 6.350                     \\
\hline\hline
\multicolumn{7}{l}{Input from \citet{tsunski} [\textit{TV11}]: $R_{p}=0.084R_{\ast}$, $R_{in}=1.11$, $R_{out}$=2.32, $i_{p}$=88$^\mathrm{o}$, $\tau=0.5$, ($u_{1},u_{2}$)=(0.2925,0.3475)} \\ \hline
$i_{r}$          & $\theta$         & TV11          & SOAP3.0      & SOAP3.0(LD)     & TV11                        & SOAP3.0                   \\ \hline
78               & 20               & 10500         & 10800        & 12374           & 3.40                        & 3.41                      \\ \hline
\end{tabular}
\end{table*}

\end{appendix}
   
\end{document}